\documentclass[aps,prl,reprint,amsmath,amssymb,superscriptaddress,nobibnotes,showpacs]{revtex4-1}

\usepackage{graphicx}
\usepackage{dsfont,color}

\begin{document}

\title{Quantum discord protection from amplitude damping decoherence}

\author{Jiwon Yune}
\affiliation{Center for Quantum Information, Korea Institute of Science and Technology (KIST), Seoul 136-791, Korea} 

\author{Kang-Hee Hong}
\affiliation{Department of Physics, Pohang University of Science and Technology (POSTECH), Pohang 790-784, Korea}

\author{Hyang-Tag Lim}
\affiliation{Department of Physics, Pohang University of Science and Technology (POSTECH), Pohang 790-784, Korea}
\affiliation{Present address : Institute of Quantum Electronics, ETH Zurich, Zurich CH-8093, Switzerland}

\author{Jong-Chan Lee}
\affiliation{Department of Physics, Pohang University of Science and Technology (POSTECH), Pohang 790-784, Korea}

\author{Osung Kwon}
\affiliation{Center for Quantum Information, Korea Institute of Science and Technology (KIST), Seoul 136-791, Korea} 
\affiliation{Present address : National Security Research Institute, Daejeon 305-600, Korea}

\author{Sang-Wook Han}
\affiliation{Center for Quantum Information, Korea Institute of Science and Technology (KIST), Seoul 136-791, Korea} 

\author{Yong-Su Kim}\email{yong-su.kim@kist.re.kr}
\affiliation{Center for Quantum Information, Korea Institute of Science and Technology (KIST), Seoul 136-791, Korea} 

\author{Sung Moon}
\affiliation{Center for Quantum Information, Korea Institute of Science and Technology (KIST), Seoul 136-791, Korea} 

\author{Yoon-Ho Kim}\email{yoonho72@gmail.com}
\affiliation{Department of Physics, Pohang University of Science and Technology (POSTECH), Pohang 790-784, Korea}

\date{\today}

\begin{abstract}
\noindent
Entanglement is known to be an essential resource for many quantum information processes. However, it is now known that some quantum features may be acheived with quantum discord, a generalized measure of quantum correlation. In this paper, we study how quantum discord, or more specifically, the measures of \textit{entropic} discord and \textit{geometric} discord are affected by the influence of amplitude damping decoherence. We also show that a protocol  deploying weak measurement and quantum measurement reversal can effectively protect quantum discord from amplitude damping decoherence, enabling to distribute quantum correlation between two remote parties in a noisy environment.
\end{abstract}

\maketitle

\section{I. Introduction}
\noindent 
Quantum correlations are essential resources that make various quantum informational processes possible, and quantum entanglement has been in the vanguard due to its fundamental roles in non-locality and advantages in many quantum information processing~\cite{NC,epr}. However, entanglement is not the only quantum correlation. Ollivier and Zurek proposed another type of quantum correlation, now known as quantum discord, from the perspective of information theory ~\cite{Ollivier}. Quantum discord is a measure of nonclassical correlations between two subsystems of a quantum system. The correlations arise from quantum physical effects. However, it does not necessarily require quantum entanglement. Hence, there exist separable states with non-zero discord. There have been significant efforts made to understand the operational meanings of quantum discord~\cite{Cavalcanti, Gu}, and find its applications in quantum information processing~\cite{Animesh, Dakic2, Stefano}. 

Quantum correlations, both entanglement and quantum discord, can be degraded by decoherence which is often caused by unavoidable coupling with the environment. There have been many studies that attempt to protect entanglement by tackling decoherence. For example, one can distill a highly entangled state from multiple copies of partially entangled states~\cite{Bennett96,Kwiat01,Pan03,Dong08}. Decoherence-free subspace~\cite{Lidar98,Kwiat00} and quantum Zeno effect~\cite{Maniscalco} can be also used to cope with decoherence. Recently, it has been shown that the weak measurement and its reversal measurement can effectively protect entanglement from the amplitude damping decoherence~\cite{kim12}. Many of these protocols might be suitable for protecting quantum discord, however, no quantitative research has been done to show the feasibility of the protection of quantum discord. Since quantum discord can exist without entanglement and it provides quantum advantages, protecting quantum discord can be useful for some quantum information tasks.

In this paper we revisit the original protocol that utilizes the weak measurement and quantum measurement reversal in order to supress the effect of decoherence~\cite{kim12,Lee,lee14,lim14} and investigate the protocol in terms of quantum discord. We theoretically and experimentally evaluate the effectiveness of quantum measurement reversal in protecting the amount of quantum discord. Our results ultimately verifies that {\it general} quantum correlations can be protected by the protocol. 

The remainder of the paper is organized as follows: After a brief review of quantum discord in Section II, we provide a numerical method to estimate quantum discord from a given density matrix in detail in Section III. Then, we introduce the weak measurement and quantum measurement reversal protocol as well as the simulation result on quantum discord in Section IV. The experimental setup and discussion is provided in Section V, and finally, in Section VI, we summarize our research and conclude.

\section{II. Quantum discord: the definition}
\noindent There exist variant versions of quantum discord, which will be introduced and discussed in the following subsections. 

\subsection{A. Entropic discord}
\noindent For a classical system, information entropy or the Shannon entropy measures the ignorance about a discrete random variable $X$ with possible values $\{x_1, x_2, ..., x_n\}$. If the probability mass function is defined as $P(x_i)$, then the Shannon entropy is defined as follows \cite{Ollivier, Henderson}:
\begin{equation}
H(X) = \sum_{i} P(x_i) I(x_i) = -\sum_{i} P(x_i) \log_b P(x_i) \text{,}
\end{equation}
\noindent where $I$ is the information content of $X$, and $b=2$ for bit. Using the definition of the Shannon entropy, we can find the mutual information of two random variables $A$ and $B$,
\begin{equation}
I(A:B) = H(A) + H(B) - H(A,B) \text{,}
\end{equation}
where $H(A, B)$ denotes the joint entropy of two random variables $A$ and $B$.

The quantum equivalence of information entropy and mutual information are similar to their classical counterparts. In quantum information theory, the entropy of a density matrix $\boldsymbol{\rho}$ is given by the von Neumann entropy, 
\begin{equation}
S(\boldsymbol{\rho}) = -{\rm Tr}(\boldsymbol{\rho}\log_b \boldsymbol{\rho}) \text{.}
\end{equation}
Note that, for a qubit, $b=2$ since this normalizes the maximum entropic information of a qubit to 1. For a joint density matrix $\boldsymbol{\rho}_{AB}$, the mutual information $\textit{I}(\boldsymbol{\rho}_{AB})$ shared by quantum systems $A$ and $B$ is given by the following equation:
\begin{equation}
\textit{I}(\boldsymbol{\rho}_{AB}) = S(\boldsymbol{\rho}_A) + S(\boldsymbol{\rho}_B) - S(\boldsymbol{\rho}_{AB}) \label{quan.cor.} \text{,}
\end{equation}
\noindent where $\boldsymbol{\rho}_A$($\boldsymbol{\rho}_B$) can be deduced by the partial trace ${\rm Tr}_{B(A)} \  \boldsymbol{\rho}_{AB}$. In order to get the amount of quantum discord $\textit{D}(\boldsymbol{\rho}_{AB})$, one  needs to deduct the measure of correlation in the classical limit $\textit{J}(\boldsymbol{\rho}_{AB})$ from the mutual quantum information $\textit{I}(\boldsymbol{\rho}_{AB})$:
\begin{eqnarray}
\textit{D}(\boldsymbol{\rho}_{AB}) = \textit{I}(\boldsymbol{\rho}_{AB}) - \textit{J}(\boldsymbol{\rho}_{AB}) \label{discord} \text{,} \\
\textit{J}(\boldsymbol{\rho}_{AB}) = \underset{\{B_k\}}{\text{sup}}\ \textit{I}(\boldsymbol{\rho}_{AB}|\{B_k\}) \label{clas.cor.} \text{,}
\end{eqnarray}
\noindent where $\{B_k\}$ is a measurement performed locally on the system $B$.

It is noteworthy that quantum discord is not generally symmetric under the exchange of the local system measurements. For instance, if we can perform a set of measurements $\{A_k\}$, instead of $\{B_k\}$, then we may get a different amount of quantum discord. Note that a symmetric discord has been proposed in order to ensure the symmetry~\cite{Wu}. Nonetheless, this paper follows the traditional definition of quantum discord, because the system of interest generally considers the environment that has symmetric effects on the systems $A$ and $B$. 

\subsection{B. Geometric discord}
\noindent Because we need to find the supremum of $\textit{I}(\boldsymbol{\rho}_{AB}|\{B_k\})$, the quantum discord between systems $A$ and $B$ is not trivial to calculate. In fact, except for special classes of states such as two-qubit X density matrices, there does not exist a closed form solution for quantum discord \cite{Modi, Huang}. As a consequence, one needs to implement complex numerical methods in order to calculate the amount of quantum discord which is presented in Sec. III. 

In order to overcome this problem, Dakic \textit{et al.} introduced geometric quantum discord that is based on the Hilbert-Schmidt distance between the density matrix $\boldsymbol{\rho}_{AB}$ and its closest classical state $\boldsymbol{\rho}_{AB}^{c}$, i.e., $D(\boldsymbol{\rho}_{AB}^{c})=0$~\cite{Dakic1, Luo}. Its definition is as follows:
\begin{equation}
\textit{D}_G(\boldsymbol{\rho}_{AB}) = \underset{\{B_k\}}{\text{inf}}\  ||\boldsymbol{\rho}_{AB} - \boldsymbol{\rho}_{AB}^{c}||_{1}  \label{geom.} \text{,}
\end{equation}
\noindent where $||X||_{1}$ is the Hilbert-Schmidt 1-norm, defined as $||X||_{1} = tr(\sqrt{X^{\dagger}X})$. This definition of quantum discord also requires numerical methods. However, the calculation process is much simpler and faster compared to the entropic definition of quantum discord since there is no need to perform logarithms of matrices.

There is another definition of geometric discord, based on the Hilbert-Schmidt 2-norm, 
\begin{equation}
\textit{D}_{G}^{(2)}(\boldsymbol{\rho}_{AB}) = \underset{\{B_k\}}{\text{inf}}\  ||\boldsymbol{\rho}_{AB} - \boldsymbol{\rho}_{AB}^{c}||_{2}^{2}  \text{,}
\end{equation}
\noindent where $||X||_{2} =\sqrt{tr(X^{\dagger}X)}$. However, recently it has been pointed out that this definition is not a good measure of quantum correlation, because it may increase under local reversible operations on the unmeasured subsystem \cite{Paula}. Hence, the discussion about the 2-norm definition will be omitted in this paper.

\section{III. Numerical methods for quantum discord estimation}

\noindent In this section, we provide numerical methods to calculate two different definitions of quantum discord, entropic discord and geometric discord. For simplicity, the discussion starts with two-qubit density matrix ($2 \otimes 2$ systems), and we extend the discussion further for any multi-qudit systems ($d \otimes d'$ systems). Note that the computational complexity of quantum discord is classified as NP-complete \cite{Huang}. Hence, resources required for computing quantum discord grow exponentially with the dimension of the Hilbert space. For any $d \otimes d'$ systems, we layout a numerical recipe for computing quantum discord based on the Monte Carlo sampling of the $d$- and $d'$-dimensional spaces. Our method does not search over the entire Hilber space, but it does give us reasonably close results, as we tested the integrity of the algorithms with repetitive trials of randomly generated density matrices with known analytical solutions.


\subsection{A. Entropic discord}

\noindent The estimation of entropic discord consists of two parts. One part is to calculate $\textit{I}(\boldsymbol{\rho}_{AB})$ (Eq. ($\ref{quan.cor.}$)), and it is fairly trivial. Note that $\boldsymbol{\rho}_{AB}$ is in the basis of $|i\rangle \otimes |j\rangle$ or $|ij\rangle$, where $i, j \in \{0, 1\}$. The other part is to find the supremum of the functional $\textit{J}(\boldsymbol{\rho}_{AB}|\{B_k\})$ (Eq. ($\ref{clas.cor.}$)). Equation ($\ref{clas.cor.}$) can be expanded to a more explicit form \cite{Ollivier}, 
\begin{equation}
\textit{J}(\boldsymbol{\rho}_{AB}) = \underset{\{B_k\}}{\text{sup}} (S(\boldsymbol{\rho}_A) - S(\boldsymbol{\rho}_{AB}|\{B_k\})) \text{.} \label{J}
\end{equation}
\noindent The second term in the equation is what requires a numerical approach. Let us define the second term of Eq. (\ref{J}) as a function, 
\begin{equation}
\textit{F}(\boldsymbol{\rho}_{AB}) = \underset{\{B_k\}}{\text{inf}} S(\boldsymbol{\rho}_{AB}|\{B_k\}) \label{func.} \textit{.}
\end{equation}
\noindent We are looking for the infimum of the functional because Eq.~(\ref{J}) has to be maximized. 

A qubit can have outcomes of either $|0\rangle$ or $|1\rangle$. However, any rotational transformation of $|0\rangle$ or $|1\rangle$ is a valid outcome of the measurement as well. Hence, for this calculation, we need to consider all the possible measurement basis.

We start with two orthogonal measurement bases $\boldsymbol{\Pi}_0$ and $\boldsymbol{\Pi}_1$,
\begin{equation}
\boldsymbol{\Pi}_0 = 
\left( \begin{array}{cc}
1 & 0 \\
0 & 0 \end{array} \right) \text{, }
\boldsymbol{\Pi}_0 = 
\left( \begin{array}{cc}
0 & 0 \\
0 & 1 \end{array} \right) \text{.}
\end{equation}
\noindent By a simple rotational transformation $\textbf{V}$, we can generalize the measurement outcome $\{B_k\}$.
\begin{equation}
\textbf{V}(\theta,\phi) = \frac{1}{\sqrt{2}} (\textbf{I} - i \hat{a}^{\dagger}(\theta,\phi) \boldsymbol{\sigma} ) \text{,}
\end{equation}
\begin{equation}
\textbf{B}_k^i  = \textbf{V}^{\dagger} \boldsymbol{\Pi}_i \textbf{V} \text{, } i \in \{0, 1\} \text{.}
\end{equation}
\noindent Note that $\hat{a}$ is a unit vector in the Bloch sphere representation,
\begin{equation}
\hat{a}(\theta,\phi) = \left( \begin{array}{c}
\sin \theta \  \cos \phi \\
\sin \theta \  \sin \phi \\
\cos \theta
\end{array} \right) \text{,}
\end{equation}
\noindent where $0 \le \theta \le \pi$ and $0 \le \phi \le 2\pi$. $\boldsymbol{\sigma}$ is a tensor of the Pauli matrices
\begin{equation}
\boldsymbol{\sigma} = \left( \begin{array}{c}
\boldsymbol{\sigma}_1 \\
\boldsymbol{\sigma}_2 \\
\boldsymbol{\sigma}_3
\end{array} \right) \text{,}
\end{equation}
\noindent where
\begin{eqnarray}
\boldsymbol{\sigma}_1 = \left( \begin{array}{cc}
0 & 1 \\
1 & 0
\end{array} \right) \text{, }
\boldsymbol{\sigma}_2 = \left( \begin{array}{cc}
0 & -i \\
i & 0
\end{array} \right) \text{, }
\boldsymbol{\sigma}_3 = \left( \begin{array}{cc}
1 & 0 \\
0 & -1
\end{array} \right) \text{.   \ \ }
\end{eqnarray}

Using the above relations, we can deduce $\boldsymbol{\rho}_{AB}$ for a given set of measurement $\{B_k\}$,
\begin{equation}
\boldsymbol{\rho}_{AB|\{B_k\}} = \sum_{i \in \{0, 1\}} \frac{1}{p_i} (\textbf{I} \otimes \boldsymbol{B}_k^i) \boldsymbol{\rho}_{AB} (\textbf{I} \otimes \boldsymbol{B}_k^i) \text{,}
\end{equation}
\noindent where $p_i$ is given by $p_i = {\rm Tr} \{ (\textbf{I} \otimes \boldsymbol{B}_k^i) \boldsymbol{\rho}_{AB} (\textbf{I} \otimes \boldsymbol{B}_k^i) \}$. It is now obvious that the functional $\textit{S}(\boldsymbol{\rho}_{AB}|\{B_k\})$ of Eq. ($\ref{func.}$) is a function of $\theta$ and $\phi$, and we can numerically estimate the extremum by simply searching over the spherical space, $0 \le \theta \le \pi$ and $0 \le \phi \le 2\pi$.


\subsection{B. Geometric discord}

\noindent Geometric quantum discord can be calculated in a similar manner. First, one needs to define an abtitrary zero quantum discord state for a given joint density matrix $\boldsymbol{\rho}_{AB}$. For this, let us define the reduced density matrix $\boldsymbol{\rho}_B$ given the measurement $|i\rangle$ of $A$, $i \in \{0, 1\}$, i.e. $\boldsymbol{\rho}_{B||0\rangle_A}$ and $\boldsymbol{\rho}_{B||1\rangle_A}$,
\begin{eqnarray}
\boldsymbol{\rho}_{B||0\rangle_A} = 
{\rm Tr}_A \{ \left[ \left( \begin{array}{cc}
1 & 0 \\
0 & 0 \end{array} \right) \otimes
\boldsymbol{I} \right] \boldsymbol{\rho}_{AB} \} \text{,} \\
\boldsymbol{\rho}_{B||1\rangle_A} = 
{\rm Tr}_A \{ \left[ \left( \begin{array}{cc}
0 & 0 \\
0 & 1 \end{array} \right) \otimes
\boldsymbol{I} \right] \boldsymbol{\rho}_{AB} \} \text{.}
\end{eqnarray}
\noindent Then, the zero quantum discord state $\boldsymbol{\rho}_{AB}^{c}$ can be found by using the following equation: 
\begin{equation}
\boldsymbol{\rho}_{AB}^{c} = \sum_{i \in \{0,1\}} \left( \textbf{V}^{\dagger} \boldsymbol{\Pi}_i \textbf{V} \right) \otimes \boldsymbol{\rho}_{B||i\rangle_A} \text{.}
\end{equation}
\noindent Using the relations described above, Eq. ($\ref{geom.}$) can also be calculated by searching over the same spherical space, $0 \le \theta \le \pi$ and $0 \le \phi \le 2\pi$.




\subsection{C. Discord estimation for arbitrary $d \otimes d'$ quantum systems}

\noindent Although this approach works fine, the optimization may be necessary for special cases. It is relatively easy to calculate the quantum discord of a two-qubit state, but for high dimensional qudits of $d>2$ where $d$ stands for a dimension of the quantum state, the searching process might take a very long time. One might encounter multiple local minima for a given arbitrary density matrix. Though we cannot yet rigorously prove which method of numerical estimation is the best way to deduce the quantum discord of an arbitrary quantum system, a number of numerical evaluations led us to the conclusion that the Monte Carlo sampling is sufficient for estimating the measure. Because it may be useful for calculating the quantum discord for a multi-qudit system, the general method is briefly discussed in the following.

For a qudit system, one can define the generalized Bloch sphere using the generalized Gell-Mann matrices, which are essentially the Pauli matrices equivalence of higher-dimensional extensions \cite{Berlmann}. For a qudit system of $d=N$, i.e., SU($N$), there are a total of $d^2-1$ Gell-Mann matrices. They can be classified into three groups: $\\$

i) $\frac{d(d-1)}{2}$ symmetric Gell-Mann matrices
\begin{equation} 
\Lambda_s^{jk} = |j\rangle \langle k|+|k\rangle \langle j|, 1 \le j < k \le d \text{,}
\end{equation}

ii) $\frac{d(d-1)}{2}$ antisymmetric Gell-Mann matrices
\begin{equation} 
\Lambda_a^{jk} = -i|j\rangle \langle k|+i |k\rangle \langle j|, 1 \le j < k \le d \text{,}
\end{equation}

iii) $(d-1)$ diagonal Gell-Mann matrices
\begin{equation} 
\Lambda_d^{l} = \sqrt{\frac{2}{l(l+1)}} \left(\sum_{j=1}^{l} |j\rangle \langle j|+ l|l+1\rangle \langle l+1| \right), \\
1 \le l \le d-1 \text{.}
\end{equation}

\noindent Using the Gell-Mann matrices, we can define the generalized Bloch vector expansion of a density matrix
\begin{equation}
\textbf{V} = \frac{1}{d} (\textbf{I}+\sqrt{d}~\vec{b} \cdot \boldsymbol{\Lambda}) \text{,}
\end{equation}
\noindent where the Bloch vector $\vec{b} = (\{b^{jk}_s\}, \{b^{jk}_a\}, \{b^l_d\})$ and $\boldsymbol{\Lambda}$ is the tensor of the generalized Gell-Mann matrices \cite{Berlmann}. Let $\Omega_d$ be the set of all points $\vec{b} \in \mathbb{R}^{d^2-1}$ such that $V$ is positive semidefinite. By definition, $\Omega \in \mathbb{R}^{d^2-1}$ is the state space or the generalized Bloch sphere. If one uses a systemmatic approach to calculate quantum discord as it is described for entropic and geometric discord for two-qubit states, we can search over all the generalized Bloch sphere, of which method consumes extensive computational resources. However, for the Monte Carlo sampling, each component in $\vec{b}$ is just a random variable. Programmatically speaking, we select $d^2-1$ random variables $\{\nu_1, \nu_2, ..., \nu_{d^2-1}\}$ and one additional random variable $r$, whose absolute values are all uniformly distributed in the range from 0 to 1. With these random variables, we can construct $\vec{b}$ by the following way:
\begin{equation}
\vec{b} = \sqrt{\frac{r}{|\nu|^2}}(\{\nu_1, \nu_2, ..., \nu_{d^2-1}\}) \text{.}
\end{equation}
\noindent By having $r$, we can cover all the possible Bloch vector, $|\vec{b}| \le 1$. Note that these randomly chosen density matrices must have physical values, i.e., they must be Hermitian and positive semidefinite. For instance, there is a possibility that diagonal terms of $\textbf{V}$ can be negative, if we carelessly applied the construction described above. One must be careful and eliminate such cases for calculation. The estimation method of the Monte Carlo sampling is tested under various cases, including two-qubit, two-qutrit cases, and any combinations of arbitrary \textit{d}-dimensional quantum systems. Using this method with a sufficiently large number of sampling can provide you a good estimation of quantum discord quickly. The plots and figures of the papers are generated with  the methods described in this section, including the Monte Carlo sampling. 

\section{IV. Theory}

\subsection{A. Weak measurement and quantum measurement reversal protocol}

\noindent Let us introduce the amplitude damping decoherence suppression protocol using the weak measurement and the quantum measurement reversal ~\cite{kim12,Lee}. Our systems of interest are two-level quantum systems (S) whose bases are $|0\rangle_{\rm S}$ and $|1\rangle_{\rm S}$. Considering an environment (E) is initially at $|0\rangle_{\rm E}$, we can model the amplitude damping decoherence~\cite{NC},
\begin{eqnarray}
|0\rangle_S \otimes |0\rangle_E  & \rightarrow  |0\rangle_S \otimes |0\rangle_E \text{,} \\
|1\rangle_S \otimes |0\rangle_E  & \rightarrow  \sqrt{\bar{D}} |1\rangle_S \otimes |0\rangle_E + \sqrt{D} |0\rangle_S \otimes |1\rangle_E \text{,}
\end{eqnarray}
\noindent where $0 \le D \le 1$ is the magnitude of the environmental decoherence and $\bar{D} \equiv 1-D$. Note that  amplitude-damping decoherence is a widely used model for various qubit systems~\cite{NC}

The experiment considers a quantum communication scenario depicted in the following. Alice prepares a two-qubit correlated state $|\Phi\rangle$,
\begin{equation}
|\Phi\rangle = \alpha |00\rangle_S + \beta|11\rangle_S \text{,}\label{input}
\end{equation}
\noindent where $|\alpha|^2+|\beta|^2=1$. This state is then delivered to Bob and Charlie through the quantum channels of which amplitude-damping decoherences are characterized as $D_1$ and $D_2$. The initially correlated state $|\Phi\rangle$ is then altered by the amplitude-damping decoherence, and the consequent two-qubit quantum state $\boldsymbol{\rho}_d$ shared by Bob and Charlie is now given as \cite{kim12}
\begin{equation}
\boldsymbol{\rho}_d =
\begin{small} \left( \begin{array}{cccc}
|\alpha|^2+D_1 D_2 |\beta|^2 & 0 & 0 & \sqrt{\bar{D}_1 \bar{D}_2} \alpha \beta^* \\
0 & D_1 \bar{D}_2 |\beta|^2 & 0 & 0 \\
0 & 0 & \bar{D}_1 D_2 |\beta|^2 & 0 \\
\sqrt{\bar{D}_1 \bar{D}_2} \alpha^* \beta & 0 & 0 & \bar{D}_1 \bar{D}_2 |\beta|^2 \end{array} \right) \end{small} \text{,}
\end{equation}
\noindent where $\bar{D}_k=1-D_k$, $k \in \{1,2\}$.

We can make it turn around by sequential operations of weak measurement ($M_{wk}$) and reversing measurement ($M_{rev}$), performed beforehand and afterward of decoherence, respectively. These operations are non-unitary and defined as follows :
\begin{eqnarray}
M_{wk} (p_1, p_2) &= \left( \begin{array}{cc}
1 & 0 \\
0 & \sqrt{1-p_1}
\end{array} \right) 
\otimes 
\left( \begin{array}{cc}
1 & 0 \\
0 & \sqrt{1-p_2}
\end{array} \right) \text{,} \\
M_{rev}(p_{r_1}, p_{r_2}) &= \left( \begin{array}{cc}
\sqrt{1-p_{r_1}} & 0 \\
0 & 1
\end{array} \right) 
\otimes 
\left( \begin{array}{cc}
\sqrt{1-p_{r_2}} & 0 \\
0 & 1
\end{array} \right) \text{,}
\end{eqnarray}
\noindent where $p_i$ and $p_{r_i}$ are the strengths of the weak measurement and the reversing measurement for Bob ($i=1$) and Charlie ($i=2$), respectively. We chose the strength for the reversing measurement for protecting the amount of correlation of the joint state $\boldsymbol{\rho}_r$ to be $p_{r_i}= (1-D_i) p_i + D_i$~\cite{kim12,Lee}. Assuming that the experiment is performing the weak and reversing measurements, the two-qubit state $\boldsymbol{\rho}_r$ is now given as
\begin{equation}
\boldsymbol{\rho}_r = \frac{1}{\textit{A}}
\begin{small} \left( \begin{array}{cccc}
|\alpha|^2+\bar{p}_1 \bar{p}_2 D_1 D_2 |\beta|^2 & 0 & 0 & \alpha \beta^* \\
0 & \bar{p}_1 D_1 |\beta|^2 & 0 & 0 \\
0 & 0 & \bar{p}_2 D_2 |\beta|^2 & 0 \\
\alpha^* \beta & 0 & 0 & |\beta|^2 \end{array} \right) \end{small} \text{,}
\end{equation}
\noindent where $\textit{A} = 1+\{\bar{p}_1 D_1(1+\bar{p}_2 D_2)+\bar{p}_2 D_2\} |\beta|^2$ and $\bar{p}_i \equiv 1-p_i$. Since we have the exact forms of the density matrices, we can analyze various quantum correlations under the amplitude damping decoherence with and without weak measurement and quantum measurement reversal protocol. Note that the entanglement behaviour has been investigated in this scenario~\cite{kim12}. The results  showed that entanglement can be protected from the amplitude damping decoherence and even entanglement sudden death phenomenon can be avoided.



\begin{figure}[t]
        \centering
	\includegraphics[width=3.4in]{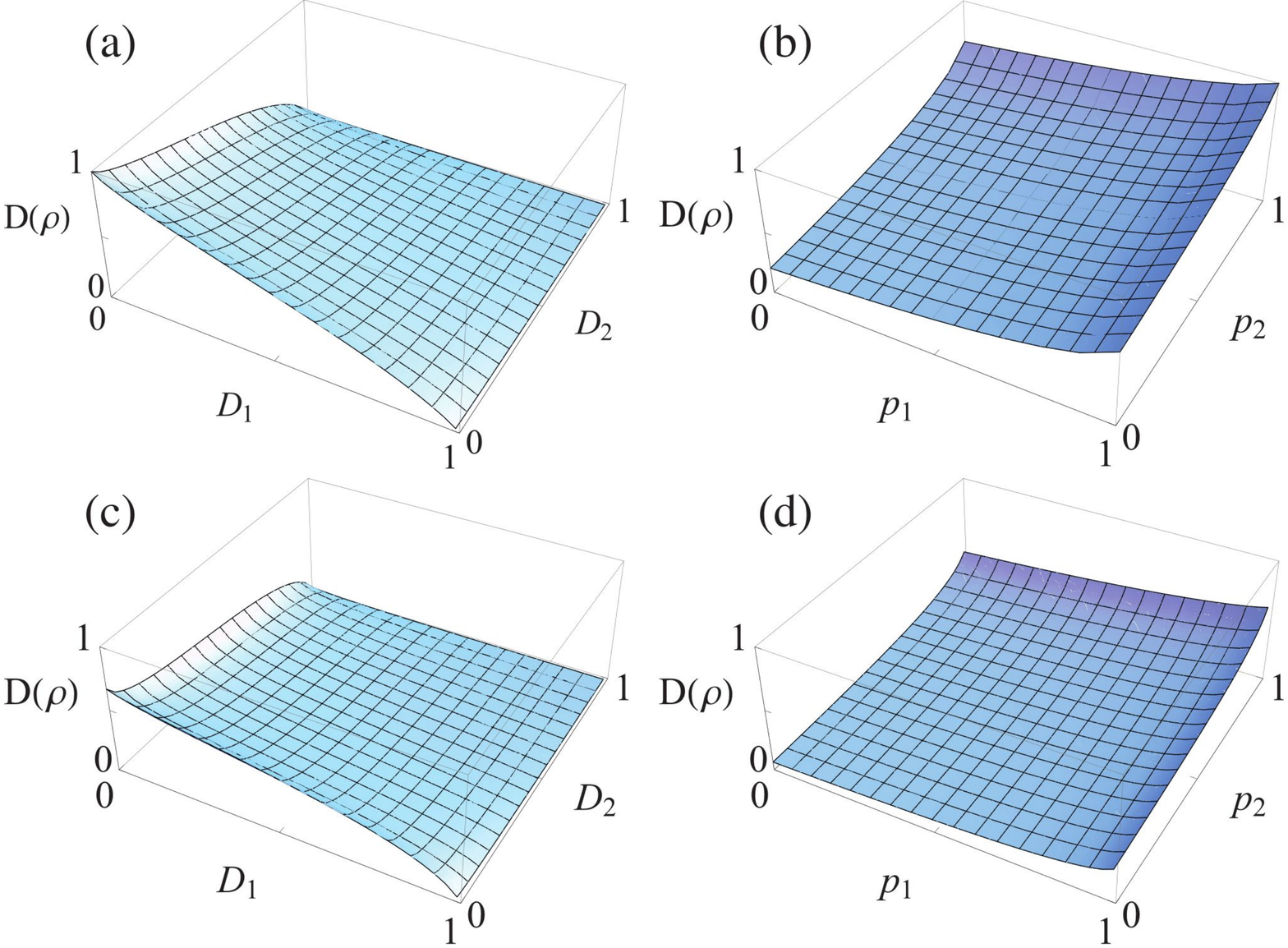}
        \caption{Theoretical estimation of entropic discord as functions of decoherence and weak measurement; Plots (a) and (b) are for the maximally correlated state $|\Phi\rangle$ with $|\alpha|=|\beta|$, and plots (c) and (d) are for the non-maximally correlated state $|\Phi\rangle$ with $|\alpha| < |\beta|$ with $\alpha=0.42$. Entropic quantum discord under the influence of decoherence is shown in plots (a) and (c), whereas the effect of the weak and reversing measurements is shown in plots (b) and (d). Plots (b) and (d) are taken with $D_1 = 0.6$ and $D_2 = 0.8$.}\label{fig:A5}
\end{figure}


\begin{figure}[b]
        \centering
	\includegraphics[width=3.4in]{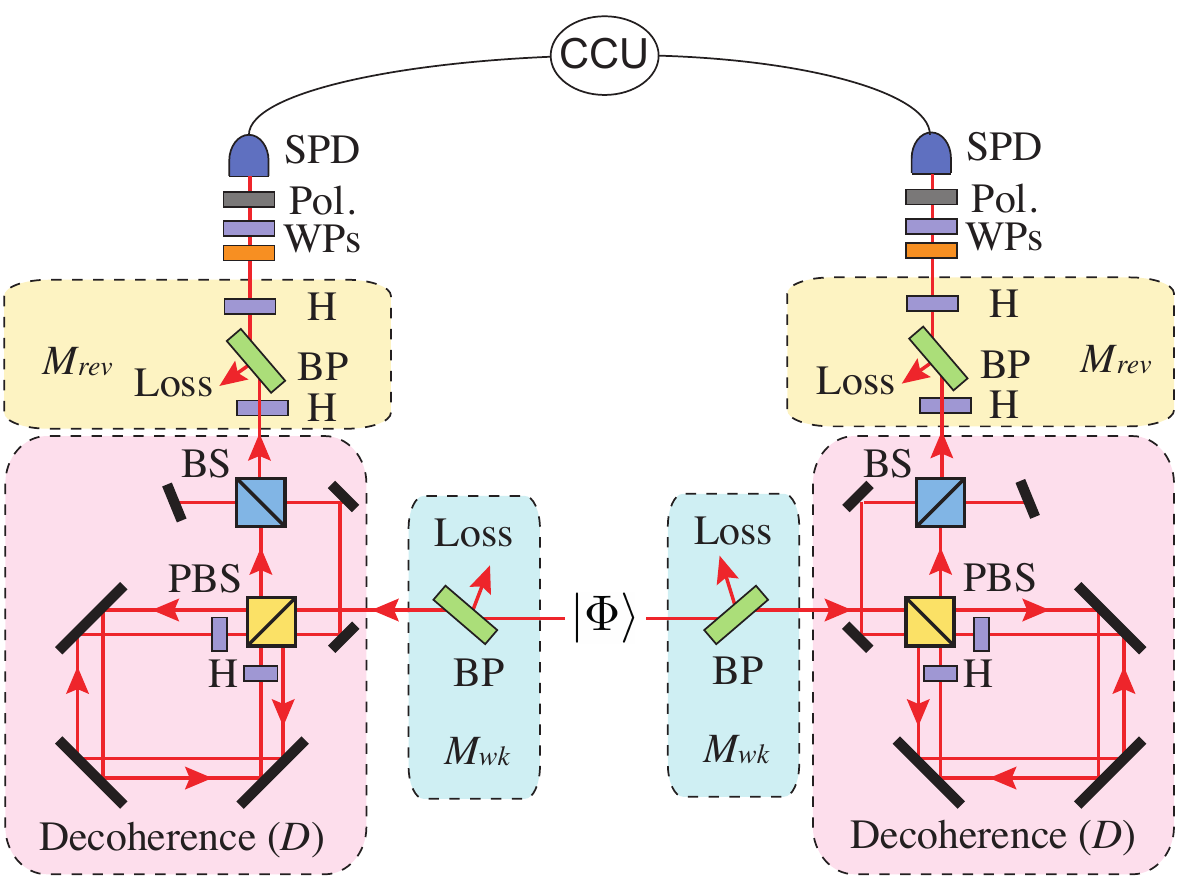}
        \caption{Experimental setup. The initial two-qubit state $|\Phi\rangle$ is the two-photon polarization state. Amplitude-damping decoherence ($D$) is implemented with the displaced Sagnac interferometers. Brewster-angle glass plates (BPs) and half-wave plates (Hs) are employed to perform weak mesurements ($M_{wk}$) and the reversing measurements ($M_{rev}$). Waveplates (WPs), polarizers (Pol.), single photon detectors (SPDs), and a coincidence counting unit (CCU) are used for quantum state tomography.}\label{setup}
\end{figure}

\subsection{Quantum discord protection}

\noindent We examine how entropic discord ($\textit{D}(\boldsymbol{\rho})$) and geometric discord ($\textit{D}_G(\boldsymbol{\rho})$) behave under different decoherence, weak measurement, and the corresponding chosen reversing measurement. Note that since both $\rho_d$ and $\rho_r$ have forms of so called X-state, there exists an analytic solution for quantum discord~\cite{Modi}. We have confirmed this analytic solution and our numerical methods in Sec. III, provide the same results. For checking the intergrity of our code, we have searched over tens of thousands randomly chosen density matrices, and it confirmed that our numerical method provides sufficiently close estimations, compared to the analytic results. Figure $\ref{fig:A5}$ shows the entropic quantum discord and the geometric quantum discord, respectively. For both cases, two particular initial states of $|\alpha|=|\beta|$ and $|\alpha|=0.42<|\beta|$ are investigated. The plots clearly show that decoherence affects the two qubits independently, and their correlations can be circumvented by exploiting weak measurement and quantum measurement reversal. However, it is noteworthy that, for quantum discord, the amplitude damping decoherence does not cause sudden death of correlation, unlike entanglement sudden death.


\begin{figure}[t]
        \centering
	\includegraphics[width=3.55in]{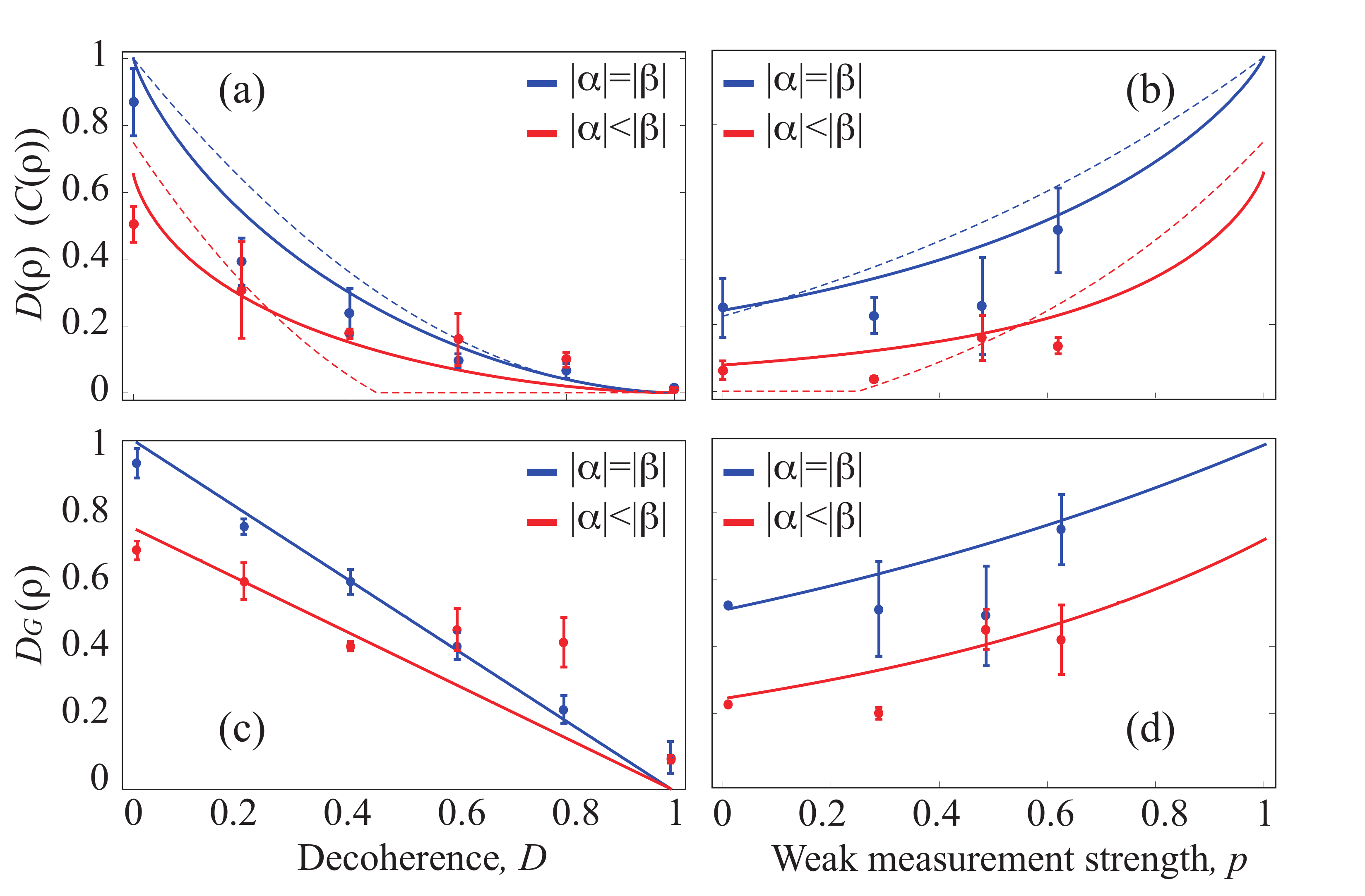}
        \caption{Experimental data for protecting \textit{entropic} (top) and \textit{geometric} (bottom) quantum discord from decoherence using quantum weak measurement and quantum measurement reversal. The blue curves are for the state $|\alpha| = |\beta|$, whereas the red ones are for $|\alpha| = 0.42$; \textbf{(a, c)} As $D$ increases, the amounts of \textit{entropic} and \textit{geometric} quantum discord gradually decrease. \textbf{(b, d)} Even under the effect of strong decoherence ($D=0.6$), we can reverse the amounts of quantum discords between Bob and Charlie by performing $M_{\text{wk}}(p)$ and $M_{\text{rev}}(p) $. The error bars represent the statistical error of $\pm 1$ standard deviation, and the dashed lines in \textbf{(a)} and \textbf{(b)} represent the corresponding concurrence plots.}\label{fig:D}
\end{figure}

\section{V. Experiment}

\noindent Figure~\ref{setup} shows the experimental setup with photonic polarization qubit implementation. First, in order to generate two-qubit entangled state, Eq.~(\ref{input}) with $|\alpha|=|\beta|$, type-I frequency-degenerate spontaneous parametric down-conversion has been implemented (not shown in Fig.~\ref{setup}). 405~nm diode laser beam is pumped into a 6-mm-thick $\beta$-BaB$_2$O$_4$ crystal to generate 810~nm photon pairs. The down-converted photons are filtered with a set of interference filters whose FWHM bandwidth is 5~nm. 

There are three main parts to implement the protocol: weak measurement, amplitude damping decoherence, and reversing measurement. The weak and reversing measurements are implemented with a set of Brewster angle glass plates (BPs) and half wave plates~\cite{kim09}. Note that because the weak and reversing measurements can be mapped to the polarization dependent losses, it is natural that the measurements can be implemented by BPs and half wave plates. 

The amplitude damping decoherence is implemented with the displaced Sagnac interferometer~\cite{Lee}. The inteferometer couples the system's polarization qubit to the environment's path qubit, whose mathematical model is provided in the beginning of Section IV. The amount of loss to the environment, or the strength of amplitude decoherence, \textit{D} can be tuned by adjusting the angle $\theta$ of the half wave plates such that  $D = \text{sin} 2\theta $.

After the protocol implementation, we perform two-photon quantum state tomography with a set of wave plates and polarizer to reconstruct the two-qubit density matrix. Note that, we have used the same experimental data of Ref.~\cite{kim12} for direct comparison between entanglement and quantum discord.

We first demonstrate the effect of decoherence $D$ on the initial two qubit mixed state $|\Phi\rangle = \alpha |00\rangle_S + \beta |11\rangle_S$. For the given $\boldsymbol{\rho}_d$, both entropic and geometric discord are evaluated. We take data points for two different input state conditions ($|\alpha|=|\beta|$ and $|\alpha|<|\beta| (|\alpha|=0.42)$) as a function of decoherence $D$, and present in Fig.~\ref{fig:D}. The dashed lines are concurrence, the amount of quantum entanglement. As observed in the figures, unless the strength of decoherence is at its maximum, i.e. $D=1$, both the entropic and geometric discords between Bob and Charlie do not disappear. This is one of the most notable difference between quantum discord and concurrence shows us that quantum discord could be a more robust resource of quantum correlation that can survive even in a severe environment than entanglement.  

We also test whether the amount of discord between Bob and Charlie can be protected by weak measurement and quantum measurement reversal. Figure $\ref{fig:D}$(b), (d) show the entropic and geometric discords of the two-qubit state $\boldsymbol{\rho}_r$, respectively. The reversing measurement parameter $p_r=p(1-D)+D$ is chosen for a given weak measurement strength $p$. As shown in Fig.~$\ref{fig:D}$, the experimental results show that the sequential operations of weak measurement and reversing measurement can indeed protect quantum discord.


\section{VI. Conclusion}
\noindent We first provided numerical methods to find both entropic and geometric discords. By applying the methods to the quantum correlation protection protocol, we successfully show that quantum discord can be protected from decoherence by weak measurement and quantum measurement reversal. The protocol described in this paper can be applied to other types of quantum system beyond two-photon polarization qubits. We believe that this protocol is a compelling method that can be used for effectively handling decoherence and distilling quantum correlations from decohered quantum resources. 

\section*{Acknowledgements}
This work was supported by the ICT R\&D programs of MSIP/IITP$[$10044559$, $2014-044-014-002$]$, the KIST Research Programs (2E25460, 2V04260, 2V04280), and the National Research Foundation of Korea (Grant No. 2013R1A2A1A01006029).

\end{document}